\documentclass[journal]{IEEEtran}
\pdfoutput=1
\usepackage{ifpdf,cite,epsf,psfrag,epsfig}
\usepackage[cmex10]{amsmath}
\usepackage{mathrsfs, amsthm,amsfonts, latexsym, amssymb,algorithmic,algorithm}
\usepackage[mathscr]{eucal}
\usepackage[caption=false,font=normalsize,labelfont=sf,textfont=sf]{subfig}
\usepackage{multicol,multirow,color}
\usepackage[breaklinks]{hyperref}
\usepackage[hyphenbreaks]{breakurl}
\usepackage{makecell}

\usepackage{array}

\begin{document}
\title{Designing model predictive control strategies for grid-interactive water heaters for load shifting applications}

\author{Elizabeth Buechler, Aaron Goldin, and Ram Rajagopal
\thanks{This work was supported in part by DOE ARPA-E under award DE-AR0000697, the Stanford Bits and Watts Initiative, a Stanford Graduate Fellowship, and a NSF Graduate Research Fellowship.}
\thanks{E. Buechler is with the Department of Mechanical Engineering, Stanford University, CA, 94305, USA (email: ebuech@stanford.edu).}
\thanks{A. Goldin is with the Department of Civil and Environmental Engineering, Stanford University, CA, 94305, USA (e-mail: agoldin@stanford.edu).}
\thanks{R. Rajagopal is with the Department of Civil and Environmental Engineering and the Department of Electrical Engineering, Stanford University, CA, 94305, USA (e-mail: ramr@stanford.edu).}
}

\maketitle
\begin{abstract}
Model predictive control (MPC) strategies allow residential water heaters to shift load in response to dynamic price signals. Crucially, the performance of such strategies is sensitive to various algorithm design choices. In this work, we develop a framework for implementing model predictive controls on residential water heaters for load shifting applications. We use this framework to analyze how four different design factors affect control performance and thermal comfort: (i) control model fidelity, (ii) temperature sensor configuration, (iii) water draw estimation methodology, and (iv) water draw forecasting methodology. We propose new methods for estimating water draw patterns without the use of a flow meter. MPC strategies are compared under two different time-varying price signals through simulations using a high-fidelity tank model and real-world draw data. Results show that control model fidelity and the number of temperature sensors have the largest impact on electricity costs, while the water draw forecasting methodology has a significant impact on thermal comfort and the frequency of runout events. Results provide practical insight into effective MPC design for water heaters in home energy management systems.
\end{abstract}



\section{Introduction}\label{sec:intro}

Load flexibility is important for the effective integration of variable renewable energy resources into electric power systems \cite{zhou2021electrification}, particularly with increasing electrification. Residential electric water heaters have the ability to shift electricity consumption without affecting thermal comfort due to their thermal storage capacity. Currently, around 46\% of residential water heaters in the United States are electric \cite{eia2020residential}, with significant load growth possible from future electrification. It is therefore essential that new electric water heaters are equipped with effective control strategies and communication capabilities to respond to grid signals and provide flexibility. 

Most water heater control strategies being implemented today in the United States use rule-based controllers designed for time-of-use (TOU) electricity rates and demand response events. For these simple use cases, rule-base strategies can be effective \cite{carew2018heat} and easily implemented with existing water heater communication standards \cite{epriCTA2045}. However in the future, water heaters may need to be able to respond to much more dynamic price signals that reflect real-time grid operating conditions, to enable more responsive demand-side flexibility \cite{madduri2022advanced}. California recently adopted rules that will require large utilities and community choice aggregators to offer an optional hourly marginal cost-based dynamic price by 2027 \cite{CPUC_rule}. To respond to dynamic signals, loads need control strategies that can anticipate future price changes and consumer water draw patterns. Strategies that have been proposed include reinforcement learning \cite{ruelens2016reinforcement,biagioni2020comparison,peirelinck2020domain}, model predictive control (MPC) \cite{jin2017foresee,jin2014model,sossan2013scheduling,shen2021data,cui2019load,blonsky2022home,biagioni2020comparison}, and other optimization-based approaches. MPC strategies are especially useful for this application as they can account for known physical dynamics and trade-offs between multiple objectives such as cost and thermal comfort. 

Developing a practically implementable MPC strategy requires making many design choices, such as selecting an optimization formulation, the locations of temperature sensors in the tank, a water draw estimation method, a water draw forecasting method, and a model parameter identification approach. Previous works primarily focus on the design of the optimization formulation for different use cases and water draw forecasting approaches. Few works have analyzed how water draw estimation accuracy or temperature sensor configuration may affect control performance. The design of these different components must be considered jointly, as their impacts on control performance and thermal comfort are often coupled. However, no publications known to the authors consider all of these components in combination. Section \ref{sec:lit_review} provides a detailed review on existing literature on MPC strategies for residential water heaters.

In this work we propose a framework for implementing model predictive controls on residential water heaters for load shifting applications. This framework addresses many of the practical challenges of implementing load shifting controls on real-world water heaters, such as having limited sensor measurements and computing power. The complete proposed framework includes an MPC optimization, water draw estimation method, water draw forecasting method, and model parameter identification method.

We use this framework to investigate how four different design choices affect control performance for two-element resistive water heaters: 
\begin{itemize}
    \item \textit{Model fidelity}: We analyze how the fidelity of the tank model used in the MPC optimization affects control performance. To do this, we extend MPC formulations developed in our previous work to a new use case. 
    \item \textit{Temperature sensor configuration}: We compare how the number of temperature sensors and their location in the tank affects control performance for each control model. This provides practical insight into how sensors should be integrated into residential water heaters.
    \item \textit{Water draw estimation}: We propose new methods for estimating water draws without a flow meter, using only the control model and temperature and power measurements, and analyze the impact of estimation errors on control performance. 
    \item \textit{Water draw forecasting}: We analyze how different water forecasting approaches affect load shifting results and thermal comfort. We propose a new approach that can be used to reduce the frequency of cold water runout events.
\end{itemize}

We observe how these design characteristics affect load shifting and thermal comfort through simulation-based testing of different strategies. These simulations use real-world water draw data and a detailed water heater model that is calibrated to laboratory measurements of a 50 gallon electric resistance water heater. Results give practical insight into how MPC design affects performance under realistic conditions, which can help enable the deployment of water heater MPC strategies in home energy management systems.

The paper is organized as follows: In Section \ref{sec:lit_review}, we review existing literature on water heater MPC strategies and define our contributions. In Section \ref{sec:water_heater_mpc}, we describe the proposed MPC control architecture and methods for draw estimation and forecasting. In Section \ref{sec:simulation_lab_testing}, the simulations used to validate control performance are described. Section \ref{sec:case_studies} shows results from case studies and Section \ref{sec:conclusions} summarizes the conclusions and potential areas for future work.

\section{Literature Review}\label{sec:lit_review}

In this section, we review the existing literature on MPC strategies for residential water heaters and identify important research gaps.

\subsection{Control model fidelity}

Water heater MPC strategies use a control model of the tank thermal dynamics. Since tank stratification patterns vary with water heater design (e.g., element/condenser coil location), the ideal MPC control model may also depend on design. Physics-based tank models vary in complexity and their ability to capture tank thermal stratification patterns. The complexity and linearity of the control model also affects the convexity and computational tractability of the optimization problem. Many previous studies use a one-node model, which neglects tank thermal stratification but results in a simple optimization problem \cite{sossan2013scheduling,shen2021data,kepplinger2016field}. Few studies analyze whether using more complex control models (e.g., constant layer volume models \cite{zinsmeister2023stratified,zuniga2017parameter} or constant layer temperature models \cite{zinsmeister2023stratified,diao2012electric}) improves control performance. We previously found that a three-node model formulation that coarsely accounts for tank stratification improves performance over a one-node model formulation for two-element resistive water heaters \cite{buechler2024improving}. Kepplinger et al. found that a multi-node model formulation improved performance over a one-node formulation for one-element resistive water heaters. Other works have analyzed how control model fidelity affects performance in space heating/cooling applications with thermal energy storage \cite{zinsmeister2023stratified,baeten2015comparison}. Additional work is needed to understand how model fidelity affects control performance for different water heater designs, particularly in combination with other MPC design choices. In this work we focus on two-element resistive water heaters, which are very common in the United States and many other countries. We compare two control models that were adapted from previous work \cite{buechler2024improving} to accommodate water heaters with thermostatic mixing valves.

\subsection{Temperature Sensor Configuration}

The impact of control model fidelity on performance also depends on how model states are measured or estimated. Particularly if simplified control models are used, the placement of temperature sensors in the tank significantly impacts plant-model mismatch, which can affect control performance. To analyze these effects accurately in simulations, high-fidelity models that accurately emulate tank stratification patterns must be used. In this work, we analyze how different practical sensor configurations affect load shifting and thermal comfort for the two different control models. To the best of our knowledge, this has not been explored in other published research.

\subsection{Estimating historical draw patterns}

Water heater MPC strategies generally use forecasts of future water use patterns in the optimization. Such forecasts can be generated from historical draw data for a specific consumer. Most previous works assume that water draws are directly measured \cite{jin2014model,shen2021data,blonsky2022home,sossan2013scheduling,cui2019load}. However, water heaters are generally not equipped with flow meters due to cost. While lower-cost vibration-based flow sensors have been explored \cite{gough2023design,pirow2018non}, sensor calibration can be sensitive to factors such as pipe material \cite{kim2008nawms}, making practical deployment challenging.

A different approach is to estimate water draws from temperature and power measurements. This has only been explored in a couple of papers. Shad et al. \cite{shad2015identification} proposed a state observer based on a one-node model. Kepplinger et al. \cite{kepplinger2015autonomous} estimated water demand by inverting a one-node tank model. Additional work is needed to understand the accuracy of different methods and how draw estimation accuracy affects MPC performance. In this work we propose a method for water draw estimation for each control model. Draw estimates are then used for forecasting future water draw patterns. Through case studies, we analyze how draw estimation accuracy affects control performance.

\subsection{Forecasting future draw patterns}

Many previous papers assume that the controller has perfect knowledge of future water draw patterns \cite{sossan2013scheduling,dela2023supervisory,amabile2021optimizing}. However, domestic hot water draw patterns are very stochastic \cite{lutz2012typical} and hard to predict. It is therefore important that control strategies perform well under large prediction errors. This requires validating control strategies using realistic water draw data. Most proposed MPC strategies are deterministic and use a single predicted water draw trajectory in the optimization. The most common approach is to forecast future draws based on historical hourly or sub-hourly average draws \cite{jin2017foresee,halamay2019hardware,knudsen2017model}. Other papers have used K-Nearest Neighbor algorithms \cite{kepplinger2015autonomous} or probabilistic approaches \cite{ritchie2021practically}. Few papers have proposed stochastic or robust optimization formulations that account for uncertainty in future water draw trajectories \cite{blonsky2022home,shen2021data}.  However, stochastic approaches can potentially be much more computationally expensive, depending on the formulation. In this work, we focus on deterministic MPC formulations and analyze how different forecasting approaches affect load shifting performance and the frequency of cold water runout events. These approaches are compared to a baseline scenario with perfect foresight of future draw patterns.

\subsection{Supervisory vs non-supervisory MPC}

MPC strategies can be categorized as either \textit{supervisory} or \textit{non-supervisory} controllers, based the control architecture. Supervisory MPC approaches \cite{dela2023supervisory} interface with a low-level controller (e.g., a thermostat) that is built into the water heater, through an API or standard communication module (e.g., CTA-2045\cite{epriCTA2045}). Non-supervisory approaches are integrated into the water heater by the manufacturer so that the controller can access all installed sensors and directly turn the elements/heat pump on and off. Supervisory strategies can more easily be implemented on existing water heaters that are equipped with an API or standardized communications module, compared to non-supervisory approaches. However, with a supervisory strategy, control actions (e.g. thermostat setpoints, loadup/shed commands) and available state information (e.g., temperatures, energy values) are limited by the low-level controller design and API. In this work, we focus on non-supervisory MPC approaches where the controller has direct access to temperature measurements and can directly turn elements on and off.

\begin{figure}
\centering
\includegraphics[width=0.47\textwidth]{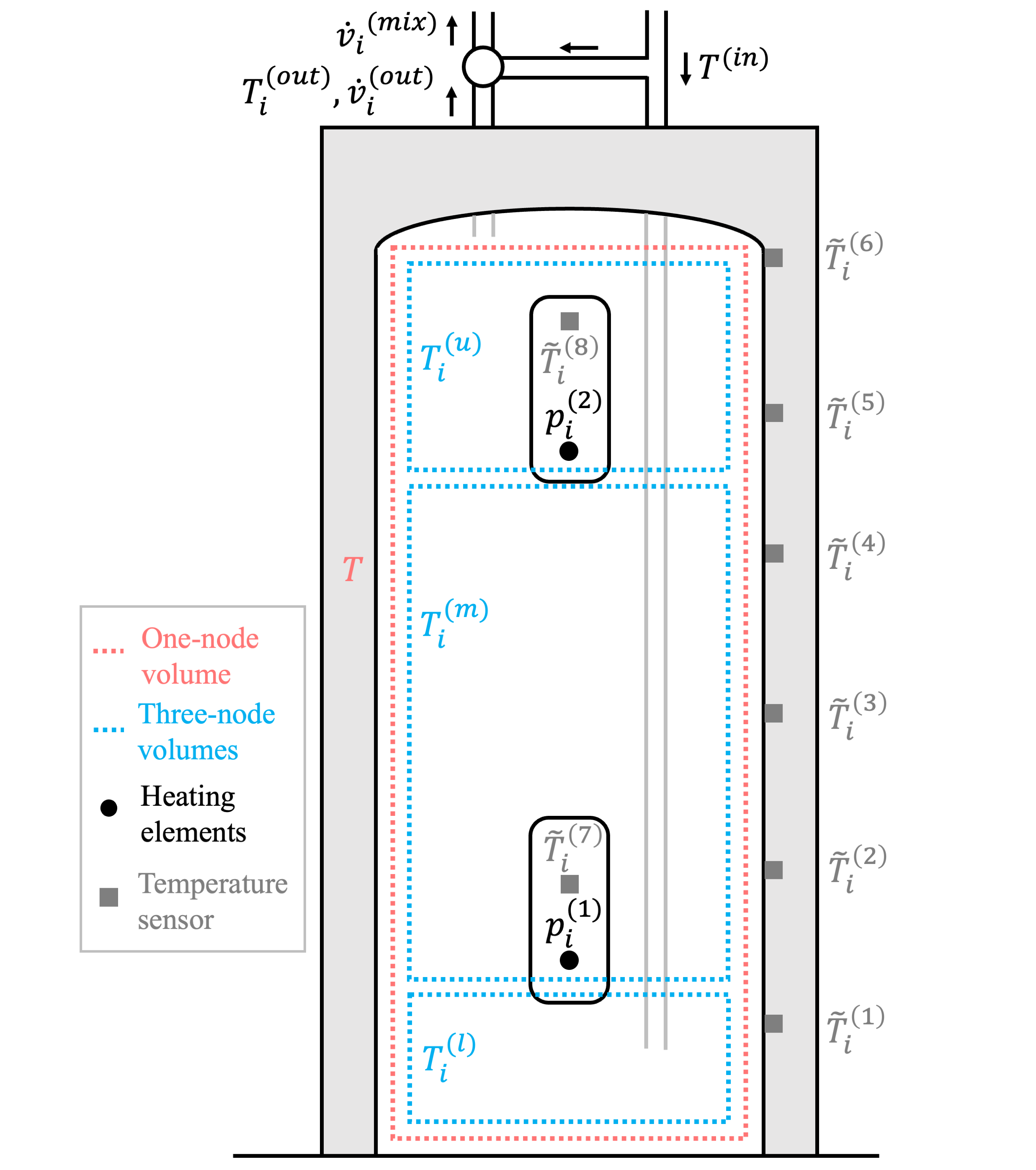}
\caption{Diagram showing the water heater temperature sensor locations, heating element locations, and volumes associated with the one-node and three-node control models.}\label{fig:model_diagram}
\end{figure}

\section{Water Heater MPC}\label{sec:water_heater_mpc}

A schematic of the modeled two-element resistive water heater under consideration is shown in Fig. \ref{fig:model_diagram}. The tank is equipped with a thermostatic mixing valve so that the water can be heated to higher temperatures without the risk of burning the consumer. The two resistive elements are wired so that they can be independently controlled. Several possible temperature sensor locations are considered, as shown in Fig. \ref{fig:model_diagram}. It is assumed that sensors would be installed on the outside of the inner metal tank under the foam insulation. Sensors 1-6 are located on the side of the water heater at approximately equal spacing, and sensors 7 and 8 are above the elements in the access bays. The MPC optimization and all other computations are run on a local low-cost microcomputer that would be either integrated into the water heater or part of a home energy management system. This microcomputer receives time-varying electricity prices from a price server (e.g., CEC MIDAS \cite{midasCEC}) using wifi or cellular communication. It also receives tank temperature and element power measurements and sends commands to turn elements on and off. 

\begin{figure*}\centering
\includegraphics[width=0.90\textwidth]{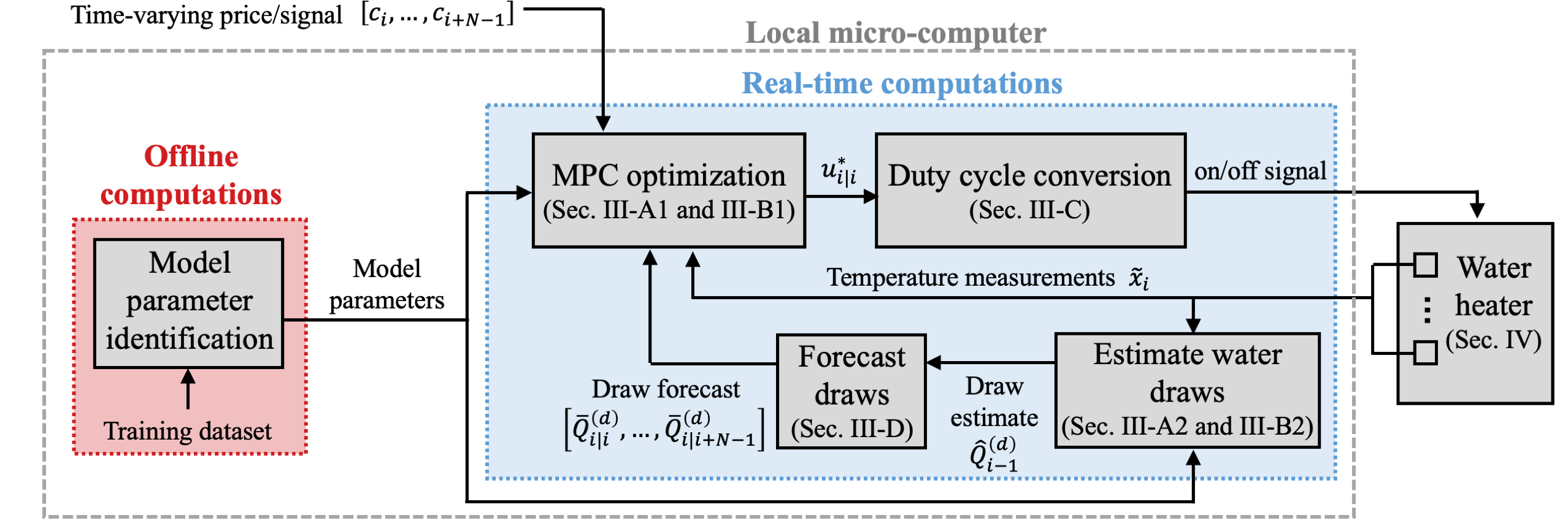}
\caption{Control architecture for water heater MPC strategies, showing all computation blocks and data streams.}\label{fig:flow_chart_large}
\end{figure*}

The objective of the proposed controller is to minimize total electricity costs under dynamic prices while maintaining thermal comfort and respecting physical constraints. The overall control architecture showing each computation block and data stream is included in Fig. \ref{fig:flow_chart_large}. Each component is summarized below and described in detail in Sections \ref{sec:one_node} and \ref{sec:3node}.

\subsubsection{MPC optimization} The dynamics of the system are optimized over a receding horizon $H$ from current timestep $\tau_i$ to $\tau_i+H$. The horizon is discretized into $N$ control intervals of length $\delta t$. The optimization variables solved at $\tau_i$ include state variables $[x_{i|i},\dots,x_{i+N|i}]$ and control variables $[u_{i|i},\dots,u_{i+N-1|i}]$, where $x_{j|i}$ and $u_{j|i}$ are the modeled state and control variables at $\tau_j$, optimized at $\tau_i$. State variables represent tank temperatures and control variables represent the average power consumption of the elements in each control interval. The optimization also uses time-varying electricity prices $[c_i,\dots,c_{i+N-1}]$ as inputs. In the optimization, the thermal dynamics of the tank are represented by a control model that defines how tank temperatures evolve as a function of element heating and water draws. In this work, we compare the performance of two different control models: a one-node model that assumes the tank is fully-mixed, and a three-node model that coarsely models tank stratification. Solving the optimization problem produces optimal control trajectory $[u^{*}_{i|i},\dots,u^{*}_{i+N-1|i}]$.

\subsubsection{Model parameter identification} The parameters of the control model used in the MPC optimization are identified via a parameter identification procedure conducted offline using a training dataset.

\subsubsection{Duty cycle conversion} The first optimal control action $u^{*}_{i|i}$ is converted from a continuous average power value to a discrete signal that is used to turn the elements on and off.

\subsubsection{Estimating and forecasting water draws} The MPC strategy also utilizes forecasts of future water consumption patterns in the optimization. During water draws, the thermostatic mixing valve mixes hot water with cold water to the mixing valve setpoint $T^{(s)}$. The volumetric flow rate out of the tank $\dot{v}^{(out)}_i$ $[m^3/s]$ is modeled as:
\begin{align}\label{eq:mix_valve_v}
    \dot{v}^{(out)}_i=\begin{cases}
    \dot{v}^{(mix)}_i\frac{(T^{(s)}-T^{(in)})}{(T_i^{(out)}-T^{(in)})} & \text{if } T^{(out)}_i\geq T^{(s)}\\
    \dot{v}^{(mix)}_i & \text{if } T_i^{(out)}< T^{(s)}
    \end{cases}
\end{align}
where $\dot{v}^{(mix)}_i$ is the volumetric flow rate of water downstream of the mixing valve and $T^{(in)}$ is the inlet water temperature $[K]$. We define $Q_i^{(d)}=\dot{v}^{(out)}_i\rho c_p (T_i^{(out)}-T^{(in)})$ as the rate at which energy is removed from the tank in drawn water $[W]$, where $c_p$ is the specific heat capacity of water $[J/kg\cdot K]$ and $\rho$ is the density of water $[kg/m^3]$. Based on Equation (\ref{eq:mix_valve_v}), $Q^{(d)}_i$ can be written as:
{\small
\begin{align}
    Q^{(d)}_i=\begin{cases}
        \dot{v}^{(mix)}_i\rho c_p (T^{(s)}-T^{(in)}) & \text{if } T^{(out)}_i\geq T^{(s)}\\
        \dot{v}^{(mix)}_i\rho c_p (T_i^{(out)}-T^{(in)}) & \text{if } T_i^{(out)}< T^{(s)}
    \end{cases}
\end{align}}
Therefore, $Q^{(d)}_i$ is not dependent on $T^{(out)}_i$ if $T^{(out)}_i\geq T^{(s)}$. Since the controller aims to keep $T^{(out)}_i\geq T^{(s)}$ to preserve thermal comfort, the optimization treats $Q^{(d)}_i$ as an exogenous variable that is not dependent on the state of the water heater. The MPC optimization solved at $\tau_i$ uses a forecast of future water draws $[\bar{Q}^{(d)}_{i|i},\dots,\bar{Q}^{(d)}_{i+N-1|i}]$ across the optimization horizon. The value $\bar{Q}^{(d)}_{j|i}$ is defined as the rate at which energy is removed from the tank in the drawn water at $\tau_j$, which is forecast at $\tau_i$. These forecasts are generated from water draw estimates from historical data over the previous $N_d$ control timesteps. At each timestep $\tau_i$, the water draw for the previous timestep $\hat{Q}^{(d)}_{i-1}$ is estimated based on measured state and control variables. This estimate is then added to a dataset of historical estimates, which is used to forecast future water draw patterns.

\subsection{One-node MPC} \label{sec:one_node}

In this section, we describe an MPC strategy based on a one-node thermal model, including the optimization formulation, water draw estimation approach, and temperature sensor configuration. \vspace{0.5em}

\subsubsection{MPC optimization}\label{sec:1node_opt}

The one-node model neglects tank thermal stratification and assumes that the tank temperature $T$ is uniform. Since the one-node model only has one state, it cannot model the separate effects of two control variables. Therefore we assume that only the bottom element is used for control. The continuous-time temperature dynamics are given by:
\begin{align}\label{eq:1node_cont}
        \frac{dT(t)}{dt}=\frac{p^{(1)}(t)}{C}+\frac{U}{C}(T^{(a)}-T(t))-\frac{Q^{(d)}(t)}{C}
\end{align}
where $C$ is the thermal capacitance of the tank $[J/K]$, such that $C=c_p \rho V$, where $V$ is the volume of the tank~$[m^3]$. $U$ is the thermal conductance of the tank insulation $[W/K]$. An explicit numerical timestepping method (e.g., forward Euler) can be used to integrate $T$ from $\tau_j$ to $\tau_{j+1}$, resulting in difference equation $T_{j+1}=f(T_j,p^{(1)}_j,Q^{(d)}_j)$.  

The following MPC optimization problem is solved at each timestep $\tau_i$. This formulation has similarities to the approach proposed in our previous work \cite{buechler2024improving}, but has been adapted for water heaters with thermostatic mixing valves. It also includes different temperature objectives and constraints that result in more effective performance.
\begingroup
\allowdisplaybreaks
\begin{subequations}
\begin{align}
    &\underset{\mathbf{T}_i,\mathbf{p}_i}{\text{minimize}}\: \sum_{j=i}^{i+N-1}{\left[\frac{\delta t}{3.6e6}c_j p^{(1)}_{j|i}+\lambda \left[T^{(min)}-T_{j|i}\right]_{+}^2\right]}\label{eq:1node_obj}\\
    &\text{subject } \text{to:}\nonumber\\  
    &\qquad T_{j+1|i}=f(T_{j|i},p^{(1)}_{j|i},\bar{Q}^{(d)}_{j|i}),\:\:\forall j\in\mathcal{J}\label{eq:1node_dyn}\\
    &\qquad T_{i|i}=\tilde{T}_{i}\label{eq:1node_init}\\
    &\qquad 0\leq p^{(1)}_{j|i}\leq p^{(max)},\:\:\forall j\in\mathcal{J}\label{eq:1node_power}\\
    &\qquad T_{j|i}\leq T^{(max)},\:\:\forall j\in\mathcal{J}\label{eq:1node_upperT}
\end{align}
\end{subequations}
\endgroup
where $\mathcal{J}=\{i,\dots,i+N-1\}$. The state variables are the tank temperatures $\mathbf{T}_i=[T_{i|i},\dots,T_{i+N|i}]$ and the control variables are the average power consumption values $\mathbf{p}_i=[p^{(1)}_{i|i},\dots,p^{(1)}_{i+N-1|i}]$. 

The objective function in Equation (\ref{eq:1node_obj}) is the weighted sum of an electricity cost term and a thermal comfort term. The thermal comfort term is positive if the tank temperature falls below $T^{(min)}$, providing a soft lower bound. This is used instead of a hard constraint to guarantee that there is always a feasible solution to the optimization problem. In our simulations, $T^{(min)}$ is set equal to the mixing valve setpoint $T^{(s)}$. Equation (\ref{eq:1node_init}) requires the initial tank temperature to be equal to the measured tank temperature $\tilde{T}_i$. Equation (\ref{eq:1node_upperT}) ensures that the modeled tank temperature stays below  $T^{(max)}$. $T^{(max)}$ may be set a few degrees below the maximum rated tank temperature. If $\tilde{T}_i$ ever exceeds $T^{(max)}$, then the MPC optimization is not run for that timestep and the elements are turned off until $\tilde{T}_i$ falls below $T^{(max)}$. Equation (\ref{eq:1node_power}) ensures that $p^{(1)}_{j|i}$, which represents the average power consumption of the lower element between $\tau_j$ and $\tau_{j+1}$, is non-negative and less than or equal to the nominal element power rating $p^{(max)}$. 

Since the objective is convex and the constraints are all affine equality or inequality constraints, the optimization problem is convex.

The parameters of the control model $V$ and $U$ are identified offline from a dataset of temperature and power measurements, as described in \cite{buechler2024improving}. This approach does not require measurements from a flow meter. \vspace{0.5em}

\subsubsection{Water draw estimation}\label{sec:1node_drawest}
We use a simple method for estimating historical water draw patterns by inverting the one-node model dynamics. The water draw from the last control period $\hat{Q}^{(d)}_{i-1}$ is estimated based on temperature measurements $\tilde{T}_{i-1}$ and $\tilde{T}_i$ and average power measurement from the previous timestep $\tilde{p}^{(1)}_{i-1}$, based on a simple energy balance. Applying forward Euler timestepping to Equation (\ref{eq:1node_cont}) and inverting the expression to solve for $\hat{Q}^{(d)}_i$ gives the following expression:

\begin{align}
    \hat{Q}^{(d)}_{i-1}=\tilde{p}^{(1)}_{i-1}+U(T^{(a)}-\tilde{T}_{i-1})-\frac{C}{\delta t}(\tilde{T}_{i}-\tilde{T}_{i-1})
\end{align}

\subsubsection{Temperature sensor configurations}\label{sec:1node_sens_config}

Particularly when there is mismatch between the control model and actual system dynamics, the positions of the sensors used to measure the state variables can significantly affect MPC performance. The one-node model assumes that the tank is fully mixed at the measured temperature, while actual tanks are stratified. For the one-node MPC, we evaluate three different temperature sensor configurations and analyze how load shifting, thermal comfort, and water draw estimation accuracy is affected. The configurations and corresponding state variable definitions are listed in Table \ref{tab:sensor_config}. The simplest configuration (1node-1) uses one temperature sensor located  above the lower element. In the two sensor configuration (1node-2s), the temperature state is equal to the average of the two sensors located above the elements (sensors 7 and 8). In the five sensor configuration (1node-5), the temperature state is equal to the average of five sensors values (sensors 2-6). Sensor 1 is not included as the volume under the lower element is not directly heated when the element is turned on. As more temperature sensors are used in these configurations, the measured state becomes a more accurate approximation of the average tank temperature.

\begin{table}[]
\caption{Temperature sensor configurations and state definitions for different MPC controllers}
    \centering
    \setlength\extrarowheight{3pt}
    \begin{tabular}{|c|c|c|}
      \hline
      \textbf{\makecell{Config.\\name }} & \textbf{\makecell{\# of\\sensors}}  & \textbf{State variable definition} \\
      \hline
        1node-1 & 1 & $\tilde{T}_i=\tilde{T}^{(7)}_i$\\
        \hline
        1node-2 & 2 & $\tilde{T}_i=(1/2)(\tilde{T}^{(7)}_i+\tilde{T}^{(8)}_i)$\\ 
        \hline
        1node-5 & 5 & $\tilde{T}_i=(1/5)\sum_{j=2}^{6}{\tilde{T}^{(j)}_i}$\\
        \hline
        & & $\tilde{T}^{(u)}_i=\tilde{T}^{(8)}_i$\\
        3node-3 & 3 & $\tilde{T}^{(m)}_i=\tilde{T}^{(7)}_i$\\
        & & $\tilde{T}^{(l)}_i=\tilde{T}^{(1)}_i$\\
        \hline
        & & $\tilde{T}^{(u)}_i=(1/2)(\tilde{T}^{(5)}_i+\tilde{T}^{(6)}_i)$\\
        3node-6 & 6 & $\tilde{T}^{(m)}_i=(1/3)(\tilde{T}^{(2)}_i+\tilde{T}^{(3)}_i+\tilde{T}^{(4)}_i)$\\
        & & $\tilde{T}^{(l)}_i=\tilde{T}^{(1)}_i$\\
       \hline
    \end{tabular}
    \label{tab:sensor_config}
\end{table}

\subsection{Three-node MPC} \label{sec:3node}

In this section we define an MPC strategy based on a three-node tank model, including the optimization formulation, water draw estimation method, and sensor configuration.\vspace{0.5em}

\subsubsection{MPC optimization}\label{sec:3node_opt}

In previous work we proposed a three-node control model that coarsely approximates stratification patterns commonly observed in two-element resistive water heaters \cite{buechler2024improving}. However, the proposed formulation did not apply to water heaters with thermostatic mixing valves. In this work, we extend the formulation to allow for thermostatic mixing valves and adapt some constraints in the formulation that have been shown to improve performance. As shown in Fig. \ref{fig:model_diagram}, the three nodes represent the volumes above the upper element, between the two elements, and below the lower element. The continuous-time thermal dynamics, which describe the temperature dynamics of the upper node $T^{(u)}_j$, middle node $T^{(m)}_j$, and lower node $T^{(l)}_j$ can be found in our previous work \cite{buechler2024improving}. 

Due to the thermostatic mixing valve, the volumetric flow rate out of the tank is modeled as:
\begin{align}\label{eq:3node_mix_valve}
    \dot{v}^{(out)}_i=\frac{Q^{(d)}_i}{\rho c_p(T^{(u)}_i-T^{(in)})}
\end{align}
which is accurate when $T^{(u)}\geq T^{(s)}$. Substituting Equation (\ref{eq:3node_mix_valve}) into the continuous-time dynamics and integrating the equations using an explicit numerical timestepping method (e.g., forward Euler) results in difference equations that are used as constraints the optimization.

The following optimization problem is solved at each timestep $\tau_i$:
{\small
\begingroup
\allowdisplaybreaks
\begin{subequations}
\begin{align}
&\underset{\substack{\mathbf{T}_{li},\mathbf{T}_{mi},\mathbf{T}_{ui}\\\mathbf{p}_{1i},\mathbf{p}_{2i}}}{\text{minimize}} \sum_{j=i}^{i+N-1}{\left[\frac{\delta t\, c_j}{3.6e6}(p^{(1)}_{j|i}+p^{(2)}_{j|i})+\lambda \left[\underline{T}-T^{(u)}_{j|i}\right]_{+}^2\right]}\\
    &\text{subject to:}\nonumber\\
    & T^{(u)}_{j+1|i}=f_u(T^{(u)}_{j|i}, T^{(m)}_{j|i},p^{(2)}_{j|i},\bar{Q}^{(d)}_{j|i}),\:\forall j\in\mathcal{J}\label{eq:3node_dyn1}\\
    &T^{(m)}_{j+1|i}=f_m(T^{(u)}_{j|i}, T^{(m)}_{j|i},T^{(l)}_{j|i},p^{(1)}_{j|i},\bar{Q}^{(d)}_{j|i}),\:\forall j\in\mathcal{J}\label{eq:3node_dyn2}\\
    &T^{(l)}_{j+1|i}=f_l(T^{(m)}_{j|i}, T^{(l)}_{j|i},\bar{Q}^{(d)}_{j|i}),\: \forall j \in \mathcal{J}\label{eq:3node_dyn3}\\
    &T^{(l)}_{j|i}\leq T^{(m)}_{j|i},\:\forall j \in \mathcal{K}\label{eq:3node_buo1}\\
    &T^{(m)}_{j|i}\leq T^{(u)}_{j|i},\: \forall j \in \mathcal{K}\label{eq:3node_buo2}\\
    &0\leq p^{(1)}_{j|i}\leq p^{(max)},\: \forall j\in\mathcal{J}\label{eq:3node_pow1}\\
    &0\leq p^{(2)}_{j|i}\leq p^{(max)},\: \forall j\in\mathcal{J}\label{eq:3node_pow2}\\
    &T^{(u)}_{i|i}=\tilde{T}^{(u)}_i &\label{eq:3node_init1}\\
    &T^{(m)}_{i|i}=\tilde{T}^{(m)}_{i} &\label{eq:3node_init2}\\
    &T^{(l)}_{i|i}=\tilde{T}^{(l)}_i &\label{eq:3node_init3}\\
    &T^{(u)}_{j|i}\leq T^{(max)},\:\forall j\in\mathcal{K}&\label{eq:3node_maxtemp}
\end{align}
\end{subequations}
\endgroup}

where $\mathcal{J}=\{i,\dots,i+N-1\}$ and $\mathcal{K}=\{i,\dots,i+N\}$. The state variables include temperature trajectories $\mathbf{T}_{xi}=[T_{i|i}^{(x)},\dots,T^{(x)}_{i+N|i}]$ for $x\in\{l,m,u\}$, and the control variables include power trajectories $\mathbf{p}_{xi}=[p_{i|i}^{(x)},\dots,p_{i+N-1|i}^{(x)}]$ for $x\in\{1,2\}$. 

Similar to the one-node MPC, the objective function is the weighted sum of an electricity cost term and a temperature penalty term. However in this case, the electricity cost term is a function of the power consumption of both elements, and the temperature penalty applies only to the upper node temperature $T^{(u)}_{j|i}$. Equations (\ref{eq:3node_pow1}) and (\ref{eq:3node_pow2}) ensure that the average power consumption of each element is non-negative and less than or equal to the nominal power rating $p^{(max)}$. Equations (\ref{eq:3node_init1})-(\ref{eq:3node_init3}) define the initial conditions and Equations (\ref{eq:3node_dyn1})-(\ref{eq:3node_buo2}) define the discrete-time control model dynamics. Equation (\ref{eq:3node_maxtemp}) ensures that the outlet temperature never exceeds the maximum tank temperature rating. If $\tilde{T}^{(u)}_i$ ever exceeds $T^{(max)}$, then both elements are turned off and the MPC is not run until it falls below $T^{(max)}$. Some water heaters may have a maximum current rating that allows only one resistive element to be on at a time. In this case, an additional constraint can be added to ensure non-simultaneous element operation, even after converting to an ON/OFF signal: 
\begin{align} \label{eq:non_sim}
    p^{(1)}_{j|i}+p^{(2)}_{j|i}\leq p^{(max)},\quad \forall j\in\mathcal{J}
\end{align}
Since the outlet flow rate is dependent on the outlet temperature (Equation (\ref{eq:3node_mix_valve})), the system dynamics are nonlinear. We solve the resulting nonlinear program in Python using the CasADi package \cite{andersson2019casadi} and the IPOPT solver \cite{wachter2006implementation}. IPOPT finds a local solution of a nonlinear program and therefore requires an initial guess. We find through simulations that solutions are relatively insensitive to initial guesses, as long as they are set to physically realistic values. However initial guesses do affect computation time and the number of solver iterations for convergence. In our work, initial guesses are warm-started using the solution from the MPC problem solved at the previous timestep.

The parameters of the control model are identified using the method described in \cite{buechler2024improving} using a dataset of power and temperature measurements. The method does not require any draw measurement data, and therefore can be applied to water heaters without a flow meter.\vspace{0.5em}

\subsubsection{Water Draw Estimation}\label{sec:3node_drawest}
Similar to the one-node MPC, we estimate $\hat{Q}^{(d)}_{i-1}$ by inverting the three-node control model. Applying forward Euler timestepping to the continuous time dynamics and combining the equations for each node gives the following expression for the total energy dynamics of the tank:

{\small
\begin{align}
    \frac{dE(t)}{dt}&\approx \frac{1}{\delta t}\sum_{x\in\mathcal{X}}{C_x(T^{(x)}_{i}-T^{(x)}_{i-1})}\nonumber\\
    &=\sum_{x\in\mathcal{X}}{\left[U_x(T^{(a)}-T^{(x)}_{i-1})\right]}-Q^{(d)}_{i-1} +p^{(1)}_{i-1} +p^{(2)}_{i-1}
\end{align}}

where $E(t)$ is the total energy stored in the water in the tank $[J]$, $\mathcal{X}=\{l,m,u\}$ is the set of nodes, $U_x$ is the thermal conductance of the tank insulation associated with each node, and $C_x$ is the thermal capacitance of each node. Inverting this expression to solve for $Q^{(d)}_{i-1}$ gives the following estimation equation:

\begin{align}
    \hat{Q}^{{d}}_{i-1}=&\tilde{p}^{(1)}_{i-1}+\tilde{p}^{(2)}_{i-1}+\sum_{x\in\mathcal{X}}{\left[U_x(T^{(a)}-\tilde{T}^{(x)}_{i-1})\right]}\nonumber\\&\quad +\sum_{x\in\mathcal{X}}{\left[-\frac{C_x}{\delta t}(\tilde{T}^{(x)}_{i}-\tilde{T}^{(x)}_{i-1})\right]}
\end{align}

\subsubsection{Temperature sensor configurations}\label{sec:3node_sensconfig}

We consider two different sensor configurations for the three-node MPC. The three sensor configuration (3node-3) has one sensor per node, with two sensors located near the elements (sensors 7 and 8) and one located below the lower element (sensor 1). The six sensor configuration uses sensors 1-6, which are approximately equally spaced along the height of the water heater. The temperature of each node is calculated from the average of the sensors placed in that node volume, as defined in Table \ref{tab:sensor_config}.

\subsection{Duty cycle conversion}\label{sec:duty_cycle}

Most resistive water heaters are controlled by switching the elements on and off. While some water heaters can continuously adjust power consumption by modulating the voltage (e.g., solar diverters), they are uncommon in the United States. We therefore focus on water heaters with only ON/OFF control capabilities.

The outputs of the MPC optimization $p^{(1)*}_{i|i}$ and $p^{(2)*}_{i|i}$ are continuous variables that represent the average power consumption of the elements in the first control interval. These values are converted to an ON/OFF signal to control the elements. This avoids using integer constraints in the MPC optimization. Unlike heat pump water heaters, resistive water heaters do not have short-cycling constraints that need to be considered.

In our simulations, we assume that the two resistive elements cannot be operated simultaneously, as defined in Equation (\ref{eq:non_sim}). The duty cycle of each element is defined as $\alpha^{(1)}_i=p^{(1)*}_{i|i}/p^{(max)}$ and $\alpha^{(2)}_i=p^{(2)*}_{i|i}/p^{(max)}$ for the lower and upper elements, respectively. For the control interval starting at $\tau_i$, the lower element is turned on from $\tau_i$ to $\tau_i+\alpha^{(1)}_i\delta t$ and the upper element is turned on from $\tau_i+\alpha^{(1)} \delta t$ to $\tau_i +(\alpha^{(1)}_i+\alpha^{(2)}_i)\delta t$. This ensures that the average power consumption values between $\tau_i$ and $\tau_{i+1}$ are equal to $p^{(1)*}_{i|i}$ and $p^{(2)*}_{i|i}$ for the lower and upper elements, respectively.

\subsection{Water draw forecasting}\label{sec:draw_forecast}

The MPC optimization solved at time $\tau_i$ uses a forecast of future water draw patterns $[\bar{Q}^{(d)}_i,\dots,\bar{Q}^{(d)}_{i+N-1}]$. This forecast is generated using a lagging set of historical draw estimates $[\hat{Q}^{(d)}_{i-N_d},\dots,\hat{Q}^{(d)}_{i-1}]$ over previous time period $H_d$, where $N_d=H_d/\delta t$.  While actual draws are non-negative physical quantities, estimated values may be negative due to plant-model mismatch. To ensure that non-physical values are not used in the MPC optimization, any negative values in the generated forecast are set equal to zero. Forecasts are generated at the same time resolution as the MPC optimization. Three approaches were investigated, as described below:

\subsubsection{Perfect foresight} In this case the draw profile is perfectly forecast based on actual draw patterns. This baseline scenario gives a measure of the maximum achievable performance of the controller given no prediction errors. 

\subsubsection{Historical mean} In this case the draw profile is forecast based on historical mean draws. Forecasted draws are equal to the mean of all draws in the historical dataset that occurred at the same time of day. In our case study, we do not differentiate between weekends and weekdays, although this could be explored in future work. 

\subsubsection{Historical quantile} Instead of using mean values, the draw profile is forecast based on quantiles of historical draws. This strategy was analyzed based on the observation that underpredicting draw volumes tends to significantly affect control performance and thermal comfort while overpredicting draw volumes has much smaller impacts. Therefore, we investigate generating conservative draw profile estimates based on large quantiles (e.g., 0.6-1.0) as a heuristic to reduce the frequency of runout events.

\section{Water Heater Simulation} \label{sec:simulation_lab_testing}

In our case studies, we evaluate MPC performance using a high-fidelity multi-node model that represents the ground-truth tank thermal dynamics. The model accounts for tank stratification along the vertical axis of the water heater, by dividing the tank into $M$ layers of equal volume \cite{buechler2024improving}. The temperature dynamics of these layers are modeled via a system of ODEs, which accounts for element heating, diffusion between adjacent layers, buoyancy, and flow from water drawn out of the tank. The mixing valve behavior is simulated using Equation (\ref{eq:mix_valve_v}). The physical dimensions of the water heater and positions of the elements were selected based on an actual 50 gallon two-element Rheem Performance water heater used in prior work \cite{buechler2024improving}. The elements have a nominal power rating of 4.5 kW at 240 V and are configured to operate non-simultaneously, due to the maximum appliance current rating. The parameters of the multi-node model were manually tuned to actual measurements from the water heater over various operating conditions. For all simulations, the number of nodes was set to 20 and the simulation timestep was set to 10 seconds, which ensured accuracy, numerical stability, and computational tractability.

\section{Case Studies}\label{sec:case_studies}

\begin{table}[]
\caption{Water draw estimation and forecasting approaches evaluated in simulations}
    \centering
    \setlength\extrarowheight{4pt}
    \begin{tabular}{|c|c|c|}
      \hline
      & \textbf{Draw estimation method} & \textbf{Draw forecast method} \\
      \hline
        1 & Perfect measurement & Perfect foresight \\
        \hline
        2 & Perfect measurement & Historical mean forecast \\
        \hline
        3 & Estimate draws & Historical mean forecast\\
        \hline
        4 & Perfect measurement & Historical quantile forecast \\
        \hline
        5 & Estimate draws & Historical quantile forecast \\
        \hline
    \end{tabular}
    \label{tab:draw_methods}
\end{table}

\begin{figure}
\centering
\includegraphics[width=0.48\textwidth]{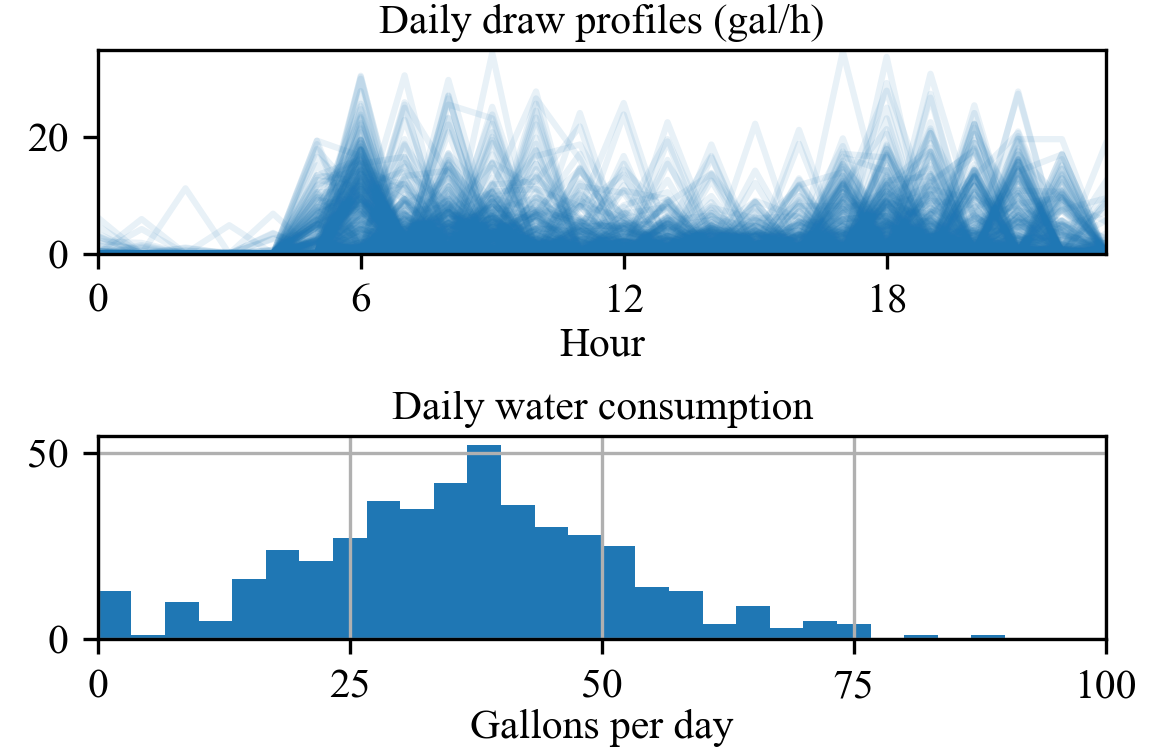}
\caption{Water draw patterns across all homes and days of simulation. Hourly draw profiles are shown on top, and a histogram of daily water consumption across all homes and days is shown on the bottom.}\label{fig:flow_patterns}
\end{figure}

Each of the MPC strategies tested through simulations is  defined by four design parameters: (i) an MPC optimization formulation, (ii) a temperature sensor configuration, (iii) a draw estimation method, and (iv) a draw forecasting method. The optimization formulations include the one-node and three-node MPC, and the corresponding temperature sensor configurations are listed in Table \ref{tab:sensor_config}. The different combinations of water draw estimation and forecasting methods that were tested for each MPC strategy are listed in Table \ref{tab:draw_methods}. This matrix of different approaches allows us to analyze the combined effects of different controller attributes on load shifting performance and thermal comfort. 

As a baseline, we also compare the performance of the MPC controllers to that of a thermostatic controller that does not perform any load shifting. The upper and lower elements are each controlled by a thermostat, with the upper and lower deadband limits set to $T^{(max)}$ and $T^{(min)}$, respectively. The elements operate non-simultaneously, with the upper element given priority.

To simulate water draw patterns, we use flow profiles published by Ritchie et al. \cite{ritchie2021practically}. The dataset includes measurements from homes over four non-consecutive one month periods. We concatenate the one-month data segments together to create a synthetic dataset with a longer duration. Eight homes were selected from the dataset that have consumption patterns that are generally reasonable for a 50 gal water heater. Fig. \ref{fig:flow_patterns} shows hourly draw profiles and a histogram of total daily water consumption across all days in the dataset for all selected homes. The dataset records the draw volume at one-minute resolution, which was used to define $\dot{v}_i^{(mix)}$. 

For the scenarios involving draw estimation, the MPC controller requires a dataset of historical draw estimates over a lagging time period $H_d$ to generate draw forecasts in real-time. To initialize the MPC controller with this historical dataset in simulation, we first ran a thermostatic controller for a time period $H_d$, during which draw estimates were generated using the proposed method. In our simulations, the thermostatic controller was run for 28 days, followed by 28 days with the MPC controller. Results were calculated on the final 26 days, excluding the first two days of MPC simulation to allow the controller to reach a steady-state behavior.

\begin{figure}
\centering
\includegraphics[width=0.48\textwidth]{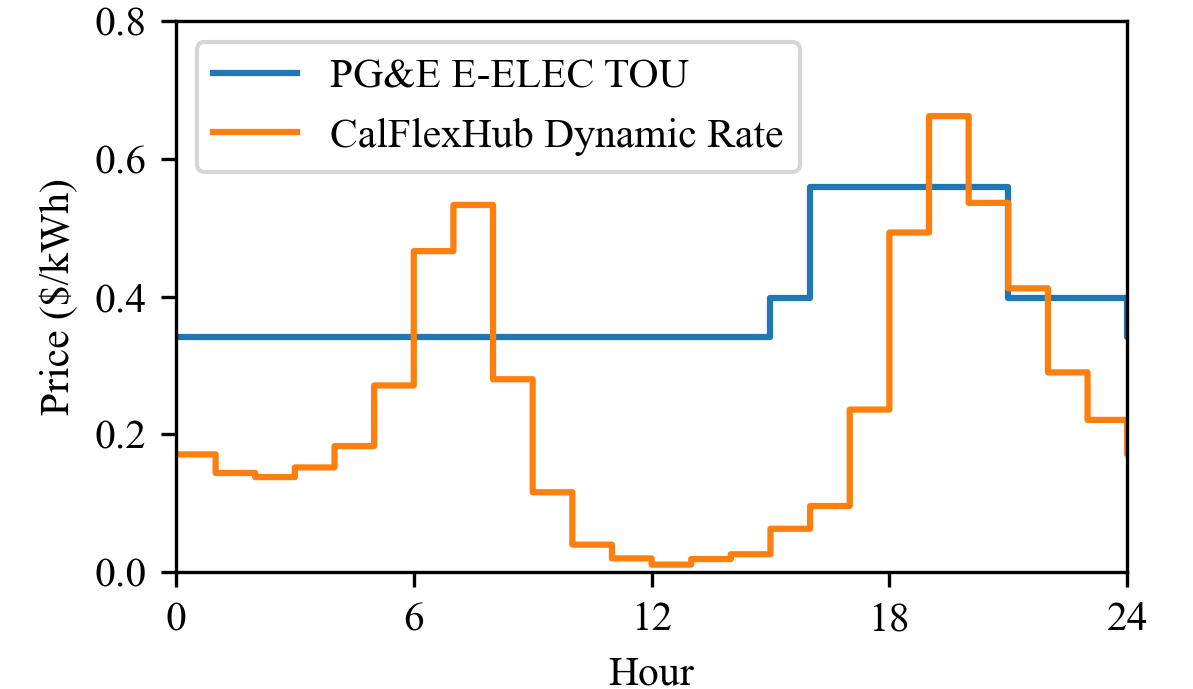}
\caption{Daily electricity price profiles that were used in simulations. The CalFlexHub Dynamic Rate is the spring HDP profile, available from the MIDAS server \cite{midasCEC}.}\label{fig:cost_profiles}
\end{figure}

The MPC strategies were evaluated on two different time-varying electricity price profiles, shown in Fig. \ref{fig:cost_profiles}: (i) a PG\&E TOU rate (E-ELEC TOU summer rate, as of July 1, 2023) and (ii) a LBNL CalFlexHub prototype dynamic retail electricity price signal. Both price signals are the same each day and known perfectly in advance. The TOU rate has a five hour peak period from 4-9pm, part-peak from 3-4pm and 9pm-12am, and off-peak from 12am-3pm. CalFlexHub has designed prototype hourly highly dynamic price (HDP) profiles that vary by season that are based off of a CalFUSE-like tariff \cite{madduri2022advanced}. These price profiles were obtained from the CEC MIDAS price server \cite{midasCEC}. In our simulations, we use the spring HDP profile, which peaks in the morning and evening and falls close to zero during the middle of the day. 

\begin{table}[]
\caption{Simulation and MPC parameters}
    \centering
    \setlength\extrarowheight{4pt}
    \begin{tabular}{|c|c|c|c|}
      \hline
       \textbf{Parameter} & \textbf{Value} & \textbf{Parameter} & \textbf{Value} \\
      \hline
        $T^{(s)}$ & 120$^\text{o}$F & $\delta t$ & 10 min \\
        \hline
        $T^{(min)}$ & 120$^\text{o}$F & $p^{(max)}$  & 4.5 kW \\
        \hline
        $T^{(max)}$ & 150$^\text{o}$F & $H$ & 24 hours \\
        \hline
        $T^{(in)}$ & 68$^\text{o}$F & $H_d$ & 28 days \\
        \hline
        $T^{(a)}$ & 70$^\text{o}$F & $M$ & 20 nodes \\
        \hline
    \end{tabular}
    \label{tab:sim_parameters}
\end{table}

Other parameter values for the simulations and the MPC controller are listed in Table \ref{tab:sim_parameters}. At the beginning of each simulation the tank was initialized to 120$^\text{o}$F. For scenarios using the historical quantile forecasting approach, we test quantiles 0.6, 0.7, 0.8 and 0.9.

\subsection{Water draw estimation accuracy}\label{sec:results_draw_est}

In this section we analyze the accuracy of the water draw estimation methods described in Section \ref{sec:one_node} and \ref{sec:3node}. Accuracy is calculated based on the root mean squared error (RMSE) of the estimates at an hourly resolution. 

\begin{figure}
\centering
\includegraphics[width=0.48\textwidth]{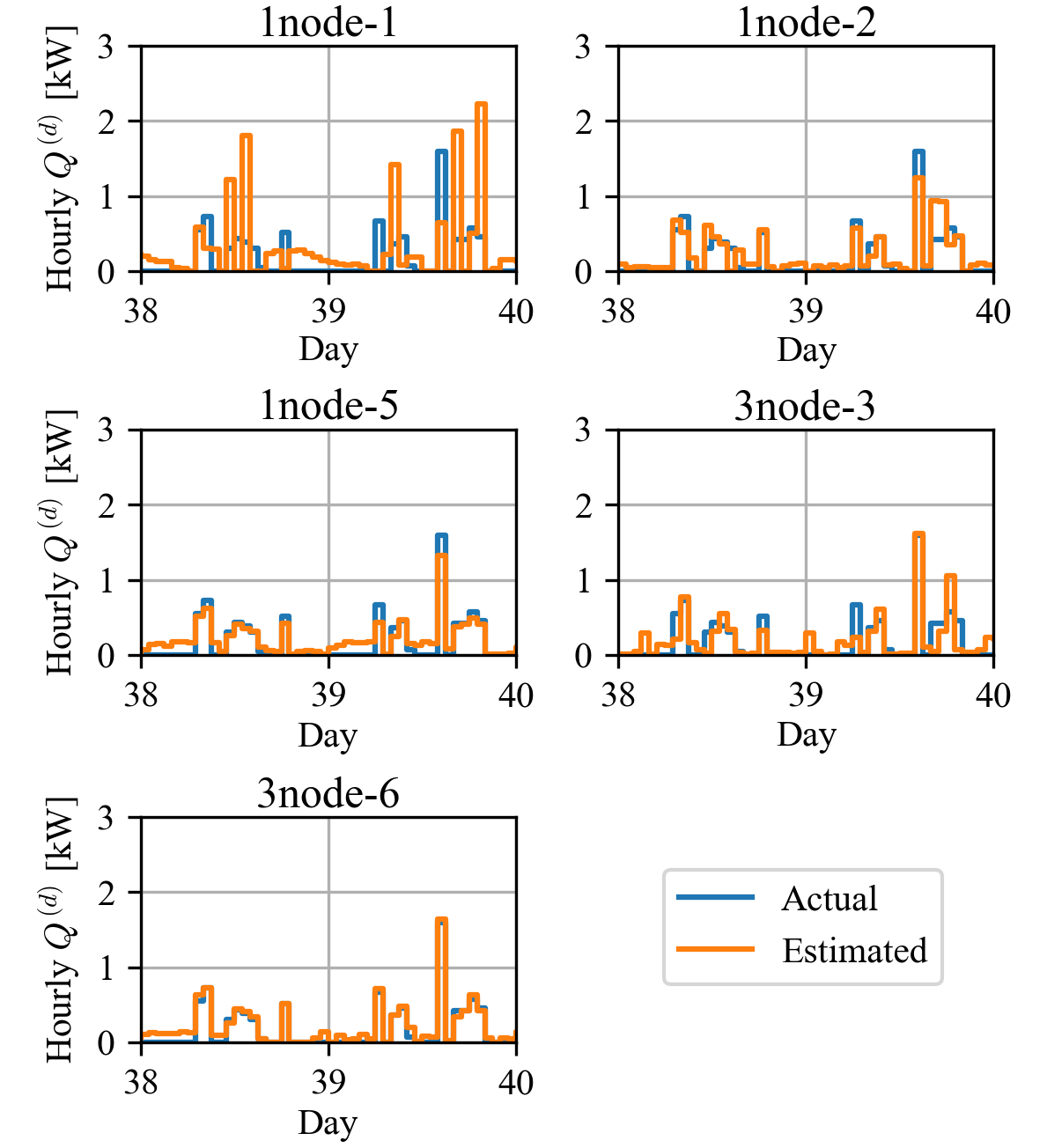}
\caption{Hourly estimated and actual water draw profiles for different control models and sensor configurations for one home over a two day period.}\label{fig:flow_est}
\end{figure}

Fig. \ref{fig:flow_est} shows actual and estimated hourly draw profiles for the different control models and sensor configurations over a two day period  for one home. Table \ref{tab:draw_rmse} shows the mean, minimum, and maximum RMSE of the estimates over all eight homes. As shown, the forecasting accuracy improves as the number of sensors increases, with the three-node six-sensor configuration obtaining the best accuracy. As shown in Fig. \ref{fig:flow_est}, the estimation errors for the six sensor scenario are mostly indiscernible, while the estimates for the one-sensor scenario deviate significantly from the actual profile.

\begin{table}[]
\caption{Mean, minimum, and maximum RMSE [kW] values of hourly water draw estimates across different homes}
    \centering
    \setlength\extrarowheight{2pt}
    \begin{tabular}{|c|c|c|c|c|}
    \hline
     \textbf{Cost}& \textbf{Controller/}& \multicolumn{3}{|c|}{\textbf{RMSE values}}\\
      \cline{3-5}
      \textbf{\makecell{profile}} & \textbf{\makecell{sensor config.}} & \textbf{Mean} & \textbf{Min} & \textbf{Max} \\
      \hline
        PG\&E & 1node-1 & 0.428 & 0.362 & 0.582 \\
        TOU    & 1node-2 & 0.193 & 0.164 & 0.280 \\
        Rate    & 1node-5 & 0.119 & 0.106 & 0.130 \\
            & 3node-3 & 0.185 & 0.168 & 0.210 \\
            & 3node-6 & 0.087 & 0.074 & 0.113 \\            
        \hline
        CalFlexHub & 1node-1 & 0.484 & 0.404 & 0.669 \\
        Dynamic    & 1node-2 & 0.208 & 0.178 & 0.301 \\
        Rate    & 1node-5 & 0.102 & 0.087 & 0.123 \\
            & 3node-3 & 0.302 & 0.248 & 0.367 \\
            & 3node-6 & 0.068 & 0.056 & 0.110 \\ 
            \hline
    \end{tabular}
    \label{tab:draw_rmse}
\end{table}

\subsection{Control performance: Load shifting}

\begin{figure*}\centering
\includegraphics[width=0.88\textwidth]{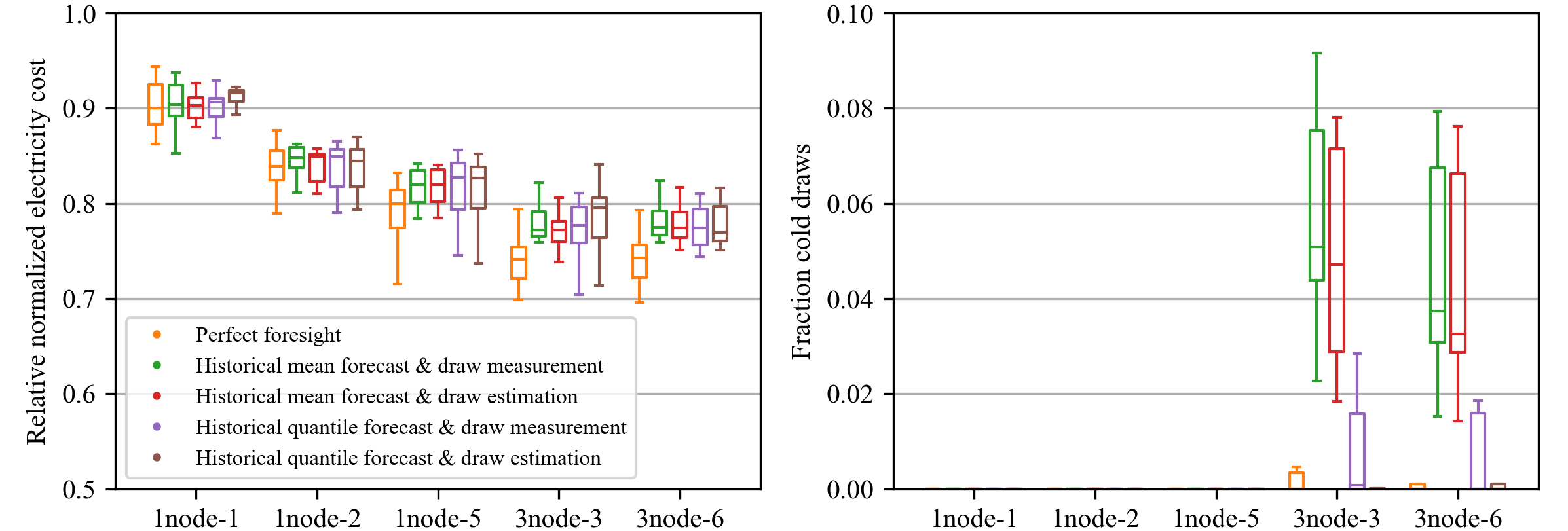}
\caption{Cost and thermal comfort results for different controllers under the PG\&E TOU price schedule. Boxplots show the distribution of results across all eight homes. To simplify the visualization, boxplot outliers are not shown. Electricity costs are relative to the thermostatic controller, and are normalized by the energy embodied in water draws. The fraction of cold draws is defined as the fraction of water draws (by volume) that were at least 10$^\text{o}$F below the mixing valve setpoint. }\label{fig:elec_tou_results}
\end{figure*}

\begin{figure*}\centering
\includegraphics[width=0.88\textwidth]{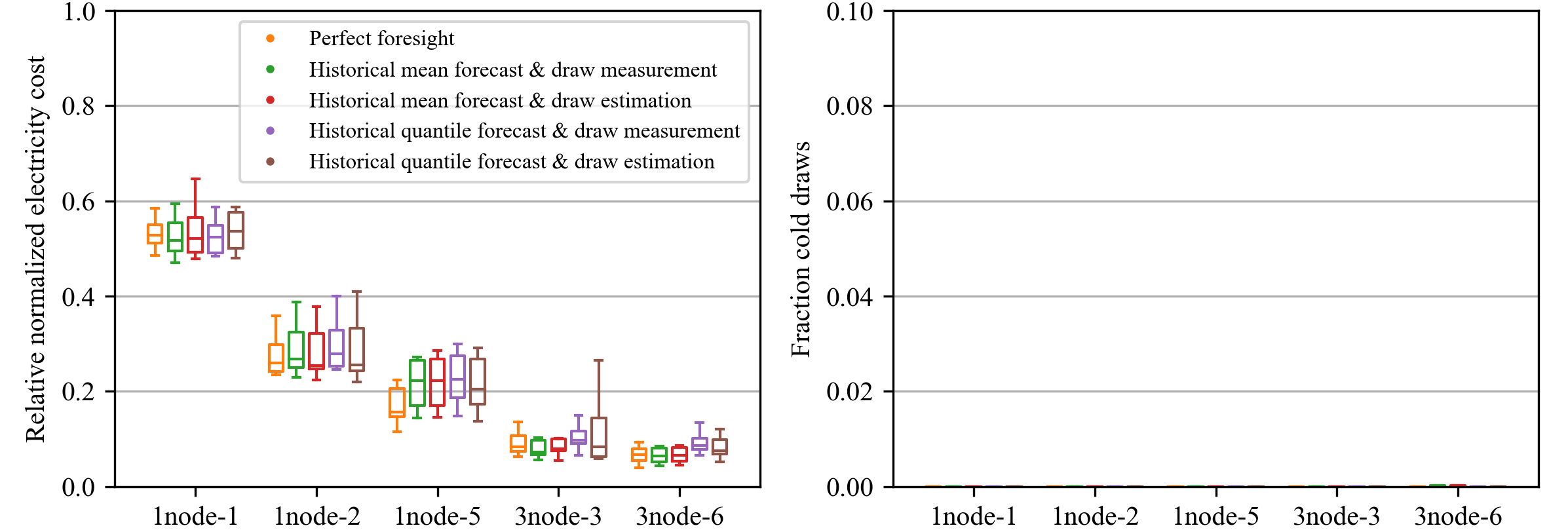}
\caption{Cost and thermal comfort results for different controllers under the CalFlexHub dynamic pricing schedule. Boxplots show the distribution of results across all eight homes. To simplify the visualization, boxplot outliers are not shown. Electricity costs are relative to the thermostatic controller, and are normalized by the energy embodied in water draws. The fraction of cold draws is defined as the fraction of water draws (by volume) that were at least 10$^\text{o}$F below the mixing valve setpoint. For the dynamic price case, cold water draws are negligible for all scenarios.}\label{fig:TPDP_results}
\end{figure*}

We quantify the load shifting performance of MPC strategies in terms of the electricity cost relative to a thermostatic controller. Electricity costs are normalized by the total energy embodied in water draws to correct for the fact that runout events can reduce electricity costs. The left-hand subplots in Figs. \ref{fig:elec_tou_results} and \ref{fig:TPDP_results} show these results for each MPC controller across all eight homes, for the TOU price profile and dynamic price profile. For the scenarios with forecasts based on historical quantiles, results are shown for the 0.9 and 0.8 quantiles for the TOU and dynamic price profiles, respectively. The impact of the choice of quantile on electricity costs and thermal comfort is discussed in the following section.

Results suggest that the control model and sensor configuration are the main factors affecting load shifting performance. The three-node MPC generally outperforms the one-node MPC, which is consistent with results from previous work \cite{buechler2024improving}. Results show the performance of the one-node MPC improves with the use of more sensors, with steady improvements between the one, two, and five sensor cases. With more sensors, the state variable becomes a better estimate of the actual average tank temperature, reducing plant-model mismatch and improving performance. Differences between sensor configurations for the three-node MPC are less significant. Overall, these results suggest that there could be benefits from manufacturers installing more temperature sensors on tanks, in terms of the resulting cost reduction and load shifting that could be obtained. Most resistive and heat pump water heaters have only two sensors, located right above the elements. The cost of including additional thermistors is negligible compared to the reduction in operating costs from improved load shifting performance over the water heater lifetime.

The electricity price profile also significantly affects load shifting and cost reduction. Cost profiles with larger ratios between maximum and minimum daily prices offer larger load shifting incentives. As shown in Figs. \ref{fig:elec_tou_results} and \ref{fig:TPDP_results}, much larger cost reductions are obtained under the dynamic price profile than the TOU profile, as the price falls nearly to zero during the middle of the day.

Results suggest that water draw estimation accuracy has a fairly small impact on load shifting performance, compared to other factors. While results in Section \ref{sec:results_draw_est} showed that estimation accuracy varies significantly with sensor configuration, costs are fairly similar between MPC strategies with perfect draw measurement and draw estimation. This suggests that installing a flow meter is not necessary for achieving good MPC performance. Historical draw patterns can be estimated with sufficient accuracy from temperature and power measurements and used to forecast future water use patterns. MPC strategies can be deployed more easily at scale if consumers do not need to install a flow meter.

\subsection{Control performance: Thermal comfort}

\begin{figure}
\centering
\includegraphics[width=0.48\textwidth]{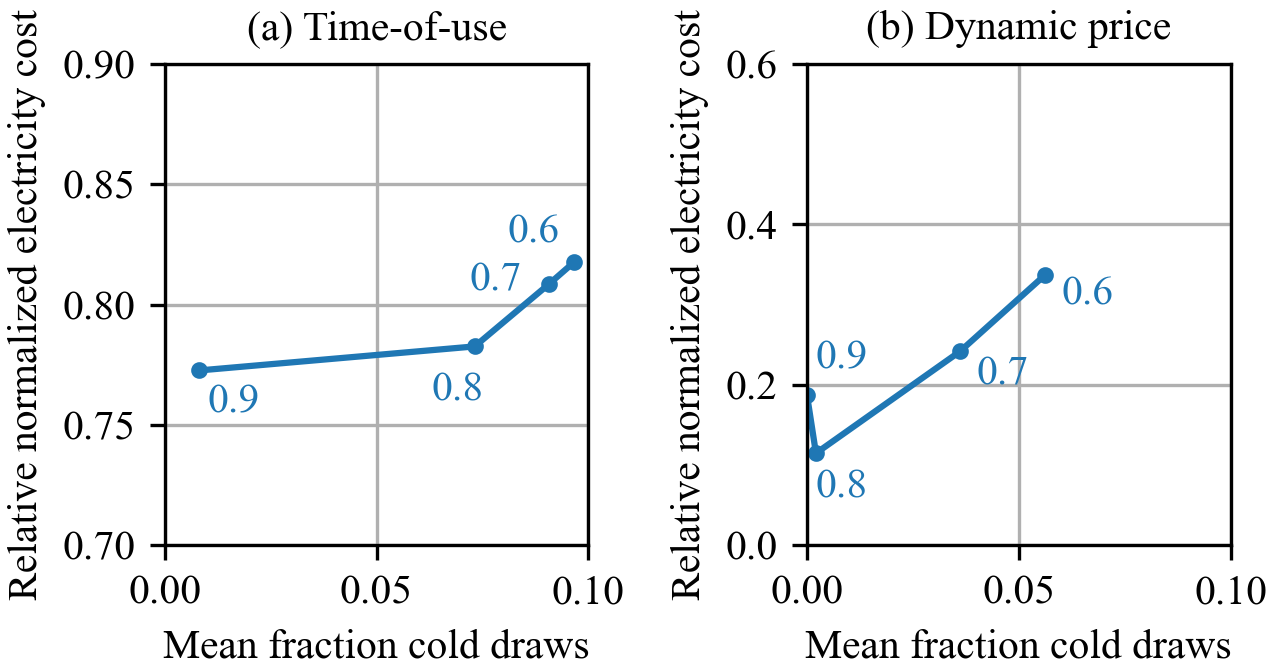}
\caption{Mean electricity cost and thermal comfort results for the three-node MPC strategy (3node-3) using the historical quantile water draw forecasting approach. Quantile values were varied between 0.6 and 0.9 and are labeled in the figure.}\label{fig:quantile_results}
\end{figure}

Thermal comfort is evaluated in terms of the fraction of water draws (by volume) that are at least 10$^\text{o}$F below the mixing valve setpoint $T^{(s)}$, as a measure of the frequency of runout events. Results show that the control model, draw forecasting approach, and the electricity price profile have the largest effect on how susceptible an MPC strategy is to runout events and thermal comfort issues. 

Results show that the three-node MPC can be more susceptible to thermal comfort issues than the one-node MPC. This occurs because the one-node MPC tends to keep the state of charge of the tank higher than the three-node MPC, due to both the design of the controller and the physics of the underlying system. Since only the bottom element can be used with the one-node MPC, most of the tank must be heated to increase the outlet temperature. With the three-node MPC, the top element can be used to heat up only the volume above the upper element. Because of this, the one-node MPC is forced to keep the tank at a higher state of charge, resulting in more conservative behavior.

However, for the three-node MPC, thermal comfort issues can be mitigated by using more conservative water draw forecasts. When water draws are forecast based on historical mean draw patterns, cold water runout events can be significant, as cumulative daily water usage is underpredicted a non-negligible fraction of the time. As shown in Fig. \ref{fig:elec_tou_results}, cold water draws occur on average around 3-5\% of the time, but can occur more frequently for certain homes. However, thermal comfort issues can be reduced almost entirely by using a more conservative draw forecast. Fig. \ref{fig:quantile_results} shows how electricity costs and the fraction of cold draws change for different quantile values when the quantile-based forecasting method is used (for the 3node-3 controller). The fraction of cold draws drops to almost zero for quantiles around 0.8-0.9. The optimal quantile that minimizes both electricity costs and the frequency of runout events depends on the electricity price profile. 

Price schedules with larger price differentials encourage more load shifting and the heating of tanks to higher temperatures, which tends to reduce the frequency of runout events. As shown in Figs. \ref{fig:elec_tou_results} and \ref{fig:TPDP_results}, runout events are negligible for the dynamic price profile, but are more common under the TOU price profile.

\subsection{Discussion}

Overall, results suggest that the three-node MPC formulation with three temperature sensors, draw estimation, and quantile-based draw forecasting achieves the best load shifting performance while also maintaining thermal comfort and requiring minimal additional sensing capabilities. Implementing this strategy would only require one additional temperature sensor, as most two-element resistive water heaters already have two sensors. It also does not require a flow meter installation. Additionally, the MPC optimization is computationally inexpensive and can be practically deployed on low-cost microcomputers. Tests of the algorithm on a Raspberry Pi unit showed an average computation time of 0.96 s for each optimization problem.

\section{Conclusions and Future Work}\label{sec:conclusions}

In this work, we performed a detailed analysis on how different aspects of MPC design affect load shifting performance and thermal comfort for residential water heaters. Four different design choices were considered: (i) the fidelity of the control model used in the optimization, (ii) the temperature sensor configuration, (iii) the water draw estimation method, and (iv) the water draw forecasting method. Results demonstrate that control model fidelity and temperature sensor resolution are very important for water heaters to be able to respond effectively to dynamic price signals. Thermal comfort depends highly on the water draw forecasting method, with conservative forecasts providing a reliable heuristic approach for reducing the frequency of runout events. Results also show that water draw patterns can be accurately estimated from historical temperature and power measurements, without requiring a flow meter. These results provide valuable practical insight into how MPC strategies should be designed for residential water heaters in home energy management systems so that they perform well under realistic scenarios.

In future work, a similar analysis could be applied to heat pump water heaters, which are becoming more widespread due to their efficiency. Heat pump water heaters have additional operating constraints and characteristics that must be accounted for in an MPC strategy. 

Future work could also analyze the performance of additional draw estimation methods and validate them through laboratory testing. The use of state estimation methods, such as Kalman filtering or state observers could potentially improve accuracy over the simple approaches proposed in this work. 

The proposed MPC strategies must be integrated into water heaters by the manufacturer, because they require direct access to sensor measurements and element controls. However, supervisory approaches that can be implemented through a standardized communications module (e.g., CTA-2045\cite{epriCTA2045}) or API could perhaps be deployed in a more scalable manner by a third-party. Future work could further investigate supervisory MPC approaches and compare the performance of supervisory and non-supervisory strategies.

\bibliographystyle{IEEEtran}
{\bibliography{ref}}
\end{document}